# Measure-resend semi-quantum private comparison without entanglement

Tian-Yu Ye*, Chong-Qiang Ye

College of Information & Electronic Engineering, Zhejiang Gongshang University, Hangzhou 310018, P.R.China

**Abstract:** In this paper, we successfully design the semi-quantum private comparison (SQPC) protocol with the measure-resend characteristic by using two-particle product states as the initial prepared quantum resource which allows two classical users to compare the equality of their private secrets under the help of a quantum third party (TP). The quantum TP is semi-honest in the sense that he is allowed to misbehave on his own but cannot conspire with either of users. Both the output correctness and the security against the outside attack and the participant attack can be guaranteed. Compared with the previous SQPC protocols, the advantage of our protocol lies in that it only employs two-particle product states as the initial prepared quantum resource, only requires TP to perform single-photon measurements and does not need quantum entanglement swapping. Our protocol can be realized with current quantum technologies.

## 1 Introduction

It is well known that quantum cryptography is an important application of quantum mechanics in the realm of cryptography. Different from classical cryptography, it theoretically gains the unconditional security by utilizing the law of quantum mechanics. Up to date, many interesting and useful branches have been derived from it, such as quantum key distribution (QKD) [1-5], quantum secure direct communication (QSDC) [6-13], quantum secret sharing (QSS) [14-18] *etc*.

Secure multi-user computation (SMC) is an important topic in classical cryptography. Accordingly, as its counterpart in the realm of quantum mechanics, quantum secure multi-user computation (QSMC) has also gained more and more attention in recent years. Quantum private comparison (QPC), as an important kind of QSMC, is devoted to accomplishing the equality comparison of secrets from two users through the transmission of quantum signals without leaking out their genuine contents. It was first put forward by Yang and Wen [19] in 2009, and has gained great developments in recent years so that numerous QPC protocols have been designed with different quantum states, such as the ones with single particles [20-22], two-particle product states [23-24], Bell states [19,21,25-33], GHZ states [34-37], W states [33,38-39], cluster states [40-41], $\chi$-type entangled states [42-44], five-particle entangled states [45], six-particle entangled states [46] and multi-level quantum system [47-48]. Lo [49] pointed out that it is impossible to design a secure equality function in a two-party scenario, hence some additional assumptions, for example, a third party (TP), are always required in QPC. After being revisited, it is easy to find out that all above QPC protocols [19-48] require all users to possess quantum capabilities. However, it sometimes may be unpractical in reality, because not all users can afford expensive quantum resources and operations.

In 2007, using the famous BB84 protocol [1], Boyer *et al.* [50] first suggested the novel concept of semi-quantumness, which means that in a quantum cryptography protocol, it is not necessary for all users to possess quantum capabilities. Boyer *et al.* [50]'s protocol is a semi-quantum key distribution (SQKD) protocol with the measure-resend characteristic. In this protocol, the receiver Bob is restricted within the following operations: (a) measuring the qubits in the fixed orthogonal basis $\{|0\rangle, |1\rangle\}$; (b) preparing the (fresh) qubits in the fixed orthogonal basis $\{|0\rangle, |1\rangle\}$; (c) sending or returning the qubits without disturbance. And the fixed orthogonal basis $\{|0\rangle, |1\rangle\}$ is thought as the classical basis, as it does not refer to any quantum superposition state. Subsequently, in 2009, Boyer *et al.* [51] designed the SQKD protocol with the randomization characteristic also using single photons. In this protocol, the receiver Bob is restricted to performing (a), (c) and (d) reordering the qubits (via different delay lines). The SQKD protocols of Refs.[50-51] are well known as the most representative pioneering works in the realm of semi-quantum cryptography. According to Refs.[50-51], the user who is only allowed to perform (a), (b), (c) and (d) is regarded to be classical.

Due to the interesting property of semi-quantumness, after invented, it was quickly absorbed into traditional QKD, QSDC, QSS, quantum key agreement (QKA), controlled deterministic secure quantum communication (CDSQC) and quantum dialogue (QD) so that SQKD [50-67], semi-quantum secure direct communication (SQSDC) [54,68], semi-quantum secret sharing (SQSS) [69-73], semi-quantum key agreement (SQKA) [74-75], controlled deterministic secure semi-quantum communication (CDSSQC) [75] and semi-quantum dialogue (SQD) [75-76] were derived, respectively. Naturally, an interesting question comes out: whether the concept of semi-quantumness can be absorbed into traditional QPC to realize semi-quantum private comparison (SQPC)? If

---

*Corresponding author:

E-mail：happyyty@aliyun.com(T.Y.Ye)

the answer to this question is positive, the equality comparison of secrets from two classical users under the help of a quantum TP may become possible. Fortunately, Chou *et al.* [77] and Thapliyala *et al.* [78] put forward two different SQPC protocols to give this question a positive answer. It can be found out that these two SQPC protocols adopt Bell entangled states as the initial prepared quantum resource and need TP to perform Bell state measurements. Moreover, the SQPC protocol of Ref.[77] employs quantum entanglement swapping.

Based on the above analysis, in order to improve the performance of the previous SQPC protocols, in this paper, we are devoted to designing the SQPC protocol with the measure-resend characteristic by just using two-particle product states as the initial prepared quantum resource. Compared with the previous SQPC protocols, the advantage of our protocol lies in that it only employs two-particle product states as the initial prepared quantum resource, only requires TP to perform single-photon measurements and does not need quantum entanglement swapping.

The rest of this paper is arranged as follows: our protocol is described in Sect.2; its output correctness and security are demonstrated in Sect.3; discussion and conclusion are given in Sect.4.

## 2 Protocol description

There are two classical users, Bob and Charlie, each of whom has one secret. Their secrets are represented by $X$ and $Y$, respectively, where $X = \sum_{j=0}^{L-1} x_j 2^j, Y = \sum_{j=0}^{L-1} y_j 2^j$ and $x_j, y_j \in \{0,1\}$. They want to determine whether $X$ and $Y$ are equal or not under the help of a quantum TP. As Yang *et al.* [29]'s semi-honest model of TP, which means that TP is allowed to misbehave on his own but will not conspire with any user, is popularly thought to be the most reasonable assumption for TP, it is naturally adopted by our protocol.

Inspired by the SQSS protocol of Ref.[71], we design the measure-resend SQPC protocol as follows.

***Preliminary:*** Bob (Charlie) divides his (her) binary representation of $X$ ($Y$) into $L$ groups $G_B^1, G_B^2, \ldots, G_B^L$ ( $G_C^1, G_C^2, \ldots, G_C^L$ ), where each group contains one binary bit.

In addition, Bob and Charlie share one key sequence $K_{BC}$ of length $L$ in advance by using the three-party circled SQKD protocol proposed by Lu and Cai in Ref.[52]. Here, $K_{BC}^i$ is the $i$ th bit of $K_{BC}$, where $K_{BC}^i \in \{0,1\}$ and $i = 1,2,\ldots,L$. Note that for clarity, Lu and Cai's three-party circled SQKD protocol is rewritten in Appendix.

**Step 1:** TP prepares $N = 8L(1+\delta)$ two-particle product states all in the state of $|++\rangle_{BC}$, where $|+\rangle = \frac{1}{\sqrt{2}}(|0\rangle + |1\rangle)$ and $\delta$ is some fixed parameter greater than 0. These two-particle product states are represented by $\{(B_1,C_1),(B_2,C_2),\cdots,(B_N,C_N)\}$, where the letters $B$ and $C$ denote two particles of each two-particle product state, and the subscripts indicate the order of two-particle product states. TP picks out particles $B$ and $C$ from each two-particle product state to form sequences $S_B$ and $S_C$, respectively. That is, $S_B = \{B_1, B_2, \cdots, B_N\}$ and $S_C = \{C_1, C_2, \cdots, C_N\}$. Finally, TP sends sequence $S_B$ to Bob and sequence $S_C$ to Charlie.

**Step 2:** When each particle arrives, Bob chooses randomly either to reflect it to TP directly (we refer to this action as CTRL), or to measure it with $\sigma_Z$ basis (i.e., the orthogonal basis $\{|0\rangle, |1\rangle\}$) and resend it to TP in the same state he found (we refer to this action as SIFT). Similarly, when each particle arrives, Charlie chooses randomly either to CTRL it or to SIFT it.

**Step 3:** TP informs Bob and Charlie of his receipt and stores the received particles in quantum memory. Bob and Charlie publish the positions of particles which they chose to CRTL.

**Step 4:** TP performs the corresponding actions on the received particles according to Bob and Charlie's choices, as indicated in Table 1.

(a) If both Bob and Charlie chose to CTRL, TP performs OPERATION 1. In this case, TP can check whether there is an Eve on the line between him and Bob or on the line between him and Charlie. If there is no Eve on the two lines, after the three participants' operations, TP should obtain $|++\rangle_{BC}$;

(b) If Bob chose to CTRL and Charlie chose to SIFT, TP performs OPERATION 2. In this case, TP can check whether there is an Eve on the line between him and Bob. If there is no Eve on the line between him and Bob, after the three participants' operations, TP should obtain $|+0\rangle_{BC}$ or $|+1\rangle_{BC}$.

(c) If Bob chose to SIFT and Charlie chose to CTRL, TP performs OPERATION 3. In this case, TP can check whether there is an Eve on the line between him and Charlie. If there is no Eve on the line between him and Charlie, after the three participants' operations, TP should obtain $|0+\rangle_{BC}$ or $|1+\rangle_{BC}$;

(d) If both Bob and Charlie chose to SIFT, TP performs OPERATION 4. The measurement result $|0\rangle$ corresponds to the classical bit 0 while the measurement result $|1\rangle$ corresponds to the classical bit 1. These classical bits are called as SIFT bits. If there is no Eve on the two lines, after the three participants' operations, their measurement results and SIFT bits should have the relations described in Table 2. Note that TP's one pair of SIFT bits corresponds to one SIFT bit from Bob and one SIFT bit from Charlie.

It is expected that approximately $\frac{N}{4}$ two-particle product states are operated by three participants in each case.

**Step 5:** TP calculates the error rates of cases (a), (b) and (c), respectively. If the error rate in any case is higher than some predefined threshold, the protocol will be terminated; otherwise, the protocol will be continued.

**Step 6:** TP calculates the error rate of case (d) as follows: He picks out at random $L$ pairs of SIFT bits to be the TEST bits and announces their positions. Then, he lets Bob and Charlie publish the values of their corresponding SIFT bits. After hearing of Bob and Charlie's announcements, he can calculate the error rate on the TEST bits by comparing the values of his pairs of SIFT bits with the values of Bob and Charlie's corresponding SIFT bits. If the error rate is higher than some predefined threshold, the protocol will be terminated; otherwise, the protocol will be continued. Also, the protocol will be aborted if there are not enough bits to perform in Step 6 or 7; this happens with exponentially small probability.

**Step 7:** For encrypting his (her) own secret, Bob (Charlie) selects at random $L$ bits from the remaining SIFT bits as the one-time pad key. Let $M_B^i$ ($M_C^i$) denote the $i$ th bit of Bob's (Charlie's) one-time pad key, where $i = 1,2,\ldots,L$. Bob (Charlie) publishes the positions of his (her) one-time pad key bits in the remaining SIFT bits. Afterward, Bob (Charlie) computes $R_B^i = G_B^i \oplus M_B^i \oplus K_{BC}^i$ ($R_C^i = G_C^i \oplus M_C^i \oplus K_{BC}^i$). Here, $\oplus$ is the modulo 2 addition operation. Finally, Bob (Charlie) publishes $R_B$ ($R_C$) to TP, where $R_B = [R_B^1, R_B^2, \cdots, R_B^L]$ ($R_C = [R_C^1, R_C^2, \cdots, R_C^L]$). Note that after hearing of the positions of Bob's (Charlie's) one-time pad key bits in the remaining SIFT bits, due to OPERATION 4 in case (d), TP can know the values of $M_B$ ($M_C$). Here, $M_B = [M_B^1, M_B^2, \cdots, M_B^L]$ ($M_C = [M_C^1, M_C^2, \cdots, M_C^L]$).

**Step 8:** For $i = 1,2,\ldots,L$: TP computes $R^i = R_B^i \oplus R_C^i \oplus M_B^i \oplus M_C^i$. If $R^i \neq 0$, TP will conclude that $X \neq Y$ and terminate the protocol. Otherwise, he will set $i = i+1$ and repeat from the beginning of this Step. If he finds out that $R^i = 0$ for all $i$ in the end, he will conclude that $X = Y$. Finally, TP tells Bob and Charlie the comparison result of $X$ and $Y$.

In our protocol, TP needs to prepare two-particle product states and perform both $\sigma_Z$ basis and $\sigma_X$ basis measurements. As a result, TP is required to possess quantum capabilities. On the other hand, both Bob and Charlie are restricted to performing the following three operations: 1) measuring the qubits in $\sigma_Z$ basis; 2) preparing the (fresh) qubits in $\sigma_Z$ basis; 3) sending or returning the qubits without disturbance. Consequently, neither Bob nor Charlie are necessary to possess quantum capabilities. It can be concluded that our protocol is indeed a SQPC protocol.

Table 1  Operations on the particles for three participants

| Case | Bob's choice | Charlie's choice | TP's operation |
| --- | --- | --- | --- |
| (a) | CTRL | CTRL | OPERATION 1 (To measure particles $B$ and $C$ with $\sigma_X$ basis, respectively) |
| (b) | CTRL | SIFT | OPERATION 2 (To measure particle $B$ with $\sigma_X$ basis and particle $C$ with $\sigma_Z$ basis) |
| (c) | SIFT | CTRL | OPERATION 3 (To measure particle $B$ with $\sigma_Z$ basis and particle $C$ with $\sigma_X$ basis) |
| (d) | SIFT | SIFT | OPERATION 4 (To measure particles $B$ and $C$ with $\sigma_Z$ basis, respectively) |

## 3  Analysis

In this section, we first demonstrate the output correctness of our protocol in Sect.3.1, then analyze its security toward the outside attack and the participant attack in Sect.3.2.

## 3.1 Output correctness

In our protocol, two classical users, Bob and Charlie, have $X = \sum_{j=0}^{L-1} x_j 2^j$ and $Y = \sum_{j=0}^{L-1} y_j 2^j$, respectively. They compare the equality of $G_B^i$ and $G_C^i$ ($i=1,2,\ldots,L$) with the help of a semi-honest quantum TP. Apparently, it can be derived that

$$
\begin{aligned}
R^i &= R_B^i \oplus R_C^i \oplus M_B^i \oplus M_C^i \\
&= \left(G_B^i \oplus M_B^i \oplus K_{BC}^i\right) \oplus \left(G_C^i \oplus M_C^i \oplus K_{BC}^i\right) \oplus M_B^i \oplus M_C^i \\
&= G_B^i \oplus G_C^i .
\end{aligned} \quad (1)
$$

According to Eq.(1), $R^i$ is the XOR value of $G_B^i$ and $G_C^i$. If $R^i = 0$, we will have $G_B^i = G_C^i$; otherwise, we will obtain $G_B^i \neq G_C^i$. It can be concluded that the output of our protocol is correct.

Table 2  Relations among three participants' measurement results and SIFT bits when both Bob and Charlie choose to SIFT

| Bob's measurement result on particle $B$ | Bob's SIFT bit corresponding to his measurement result on particle $B$ | Charlie's measurement result on particle $C$ | Charlie's SIFT bit corresponding to her measurement result on particle $C$ | TP's measurement results on particles $B$ and $C$ | TP's pair of SIFT bits corresponding to his measurement results on particles $B$ and $C$ |
|---|---|---|---|---|---|
| $\lvert 0 \rangle$ | 0 | $\lvert 0 \rangle$ | 0 | $\lvert 0 \rangle \lvert 0 \rangle$ | 00 |
| $\lvert 0 \rangle$ | 0 | $\lvert 1 \rangle$ | 1 | $\lvert 0 \rangle \lvert 1 \rangle$ | 01 |
| $\lvert 1 \rangle$ | 1 | $\lvert 0 \rangle$ | 0 | $\lvert 1 \rangle \lvert 0 \rangle$ | 10 |
| $\lvert 1 \rangle$ | 1 | $\lvert 1 \rangle$ | 1 | $\lvert 1 \rangle \lvert 1 \rangle$ | 11 |

## 3.2 Security

### 3.2.1 The outside attack

We analyze the outside attack according to each step of our protocol.

The protocol from Step 1 to Step 6 is similar to the SQSS protocol of Ref.[71]. An outside eavesdropper, Eve, may try to launch some famous attacks on the transmitted particles, such as the intercept-resend attack, the measure-resend attack, the entangle-measure attack and the Trojan horse attacks, to obtain some useful information about Bob and Charlie's SIFT bits.

The intercept-resend attack means that Eve intercepts the particles sent from TP to Bob (Charlie) and then sends her prepared fake particles to Bob (Charlie). As for this kind of attack, Eve will inevitably be detected because of two aspects: on the one hand, Eve has to randomly prepare her fake particles; on the other hand, Bob and Charlie's choices of operations are random to Eve. For example, assume that Eve happens to prepare her fake particles in the state $\lvert +0 \rangle$ and then sends $\lvert + \rangle$ and $\lvert 0 \rangle$ to Bob and Charlie, respectively. In the cases of (b) and (d) in Table 1, this kind of attack from Eve induces no error. However, in the case of (a), after Bob, Charlie and TP's operations, $\lvert +0 \rangle$ is collapsed into $\lvert ++ \rangle$ or $\lvert +- \rangle$ each with equal probability. As a result, Eve will be discovered by TP with the probability of $50\%$. In the case of (c), after Bob, Charlie and TP's operations, $\lvert +0 \rangle$ is collapsed into $\lvert 0+ \rangle, \lvert 0- \rangle, \lvert 1+ \rangle$ or $\lvert 1- \rangle$ each with equal probability. As a result, Eve will be discovered by TP with the probability of $50\%$. To sum up, the average error rate introduced by Eve in the four cases is $25\%$. It should be pointed out that if Eve happens to prepare all particles $B$ and $C$ in the state $\lvert ++ \rangle$, she will not be detected, as her fake particles are the same as the original ones. However, this situation happens only with the probability of $\left(\frac{1}{16}\right)^N$, which converges to 0 when $N$ is large enough. To say the least, even if this situation occurs, Eve still gets nothing about Bob and Charlie's SIFT bits.

The measure-resend attack means that Eve intercepts the particles sent from TP to Bob (Charlie), measures them randomly in $\sigma_Z$ basis or $\sigma_X$ basis and sends the measured states to Bob (Charlie). As for this kind of attack, Eve may be detected because of two aspects: on the one hand, Eve's measurements may destroy the original states of particles $B$ and $C$; on the other hand, Bob and Charlie's choices of operations are random to Eve. Without loss of generality, take the situation of Eve's measuring particles $B$ and $C$ in $\sigma_Z$ basis for example. Then, particles $B$ and $C$ are collapsed into $|00\rangle$, $|01\rangle$, $|10\rangle$ or $|11\rangle$ each with equal probability. Assume that particles $B$ and $C$ are collapsed into $|01\rangle$. Then, Eve sends $|0\rangle$ and $|1\rangle$ to Bob and Charlie, respectively. In the case of (d) in Table 1, this attack from Eve induces no error. However, in the case of (a), after Bob, Charlie and TP's operations, $|01\rangle$ is collapsed into $|++\rangle$, $|+-\rangle$, $|-+\rangle$ or $|--\rangle$ each with equal probability. As a result, Eve will be discovered by TP with the probability of $75\%$. In the case of (b), after Bob, Charlie and TP's operations, $|01\rangle$ is collapsed into $|+1\rangle$ or $|-1\rangle$ each with equal probability. As a result, Eve will be discovered by TP with the probability of $50\%$. In the case of (c), after Bob, Charlie and TP's operations, $|01\rangle$ is collapsed into $|0+\rangle$ or $|0-\rangle$ each with equal probability. As a result, Eve will be discovered by TP with the probability of $50\%$. To sum up, the average error rate introduced by Eve in the four cases of this situation is $43.75\%$. It should be pointed out that if Eve happens to measure all particles $B$ and $C$ in $\sigma_X$ basis, she will not be detected, as her attack makes the original states of all particles $B$ and $C$ unchanged. However, this situation happens only with the probability of $\left(\frac{1}{4}\right)^N$, which converges to 0 when $N$ is large enough. To say the least, even if this situation occurs, Eve still gets nothing about Bob and Charlie's SIFT bits.

Eve's entangle-measure attack can be modeled as two unitaries: $U_E$ attacking particles as they go from TP to Bob and Charlie and $U_F$ attacking particles as they go back from Bob and Charlie to TP, where $U_E$ and $U_F$ share a common probe space with initial state $|0\rangle_E$. As pointed out in Refs.[50-51], the shared probe allows Eve to make the attack on the returning particles depend on knowledge acquired by $U_E$ (if Eve does not take advantage of this fact, then the "shared probe" can simply be the composite system comprised of two independent probes). Any attack where Eve would make $U_F$ depend on a measurement made after applying $U_E$ can be implemented by unitaries $U_E$ and $U_F$ with controlled gates. Eve's entangle-measure attack within the implementation of the protocol is depicted in Fig.1 [71]. Ref.[71]'s Theorem 1 and Remark 2 show that the final state of Eve's probe is independent from Bob and Charlie's measurement results. However, the proof of this conclusion is not complete in Ref.[71]. For the sake of completeness, we rewrite Theorem 1 and give its whole proof as follows.

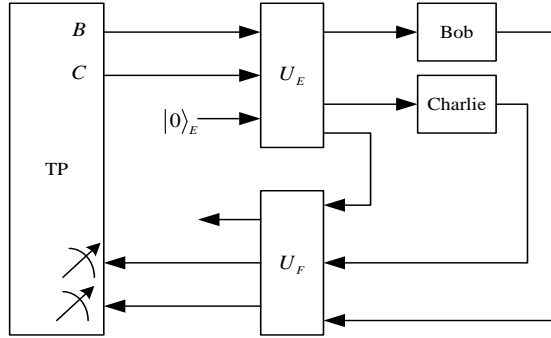

Fig 1. Eve's entangle-measure attack with two unitaries $U_E$ and $U_F$

**Theorem 1.** *Suppose that Eve performs attack $(U_E, U_F)$ on the particles from TP to Bob and Charlie and back to TP. For this attack inducing no error in Steps 5 and 6, the final state of Eve's probe should be independent of Bob and Charlie's measurement results. As a result, Eve gets no information on the SIFT bits of Bob and Charlie.*

**Proof.** Before Eve's attack, the global state of the composite system composed by particles $B$, $C$ and $E$ is $|+\rangle_B|+\rangle_C|0\rangle_E$. After Eve has performed $U_E$, the global state evolves into

$$|\psi\rangle = U_E(|+\rangle_B|+\rangle_C|0\rangle_E) = |00\rangle_{BC}|E_{00}\rangle + |01\rangle_{BC}|E_{01}\rangle + |10\rangle_{BC}|E_{10}\rangle + |11\rangle_{BC}|E_{11}\rangle, \tag{2}$$

where $|E_{ij}\rangle$s are un-normalized states of Eve's probe.

When Bob and Charlie receive the particles sent from TP, they choose either to CTRL or to SIFT. After that, Eve performs $U_F$ on the particles sent back to TP.

(i) Firstly, consider the case that both Bob and Charlie have chosen to SIFT. As a result, the state of $B+C+E$ will be $|x_1x_2\rangle_{BC}|E_{x_1x_2}\rangle$, where $x_1, x_2 \in \{0,1\}$. For Eve not being detectable in Step 6, $U_F$ should establish the following relations:

$$U_F(|x_1x_2\rangle_{BC}|E_{x_1x_2}\rangle) = |x_1x_2\rangle_{BC}|F_{x_1x_2}\rangle, \tag{3}$$

which means that $U_F$ cannot change the state of $B+C$ after Bob and Charlie's operations. Otherwise, Eve will be detected with a non-zero probability.

(ii) Secondly, consider the case that Bob has chosen to SIFT and Charlie has chosen to CRTL. As a result, the state of $B+C+E$ will be $|00\rangle_{BC}|E_{00}\rangle+|01\rangle_{BC}|E_{01}\rangle$ if Bob's measurement result is $|0\rangle$ or $|10\rangle_{BC}|E_{10}\rangle+|11\rangle_{BC}|E_{11}\rangle$ if Bob's measurement result is $|1\rangle$.

Assume that Bob's measurement result is $|0\rangle$. After Eve performs $U_F$ on the particles sent back to TP, due to Eq.(3), the state of $B+C+E$ evolves into

$$U_F\left(|00\rangle_{BC}|E_{00}\rangle+|01\rangle_{BC}|E_{01}\rangle\right)=|00\rangle_{BC}|F_{00}\rangle+|01\rangle_{BC}|F_{01}\rangle=|0\rangle_B\left(|0\rangle_C|F_{00}\rangle+|1\rangle_C|F_{01}\rangle\right). \tag{4}$$

Let $|\psi_0\rangle=|0\rangle_C|F_{00}\rangle+|1\rangle_C|F_{01}\rangle$. Replacing $|0\rangle$ with $\frac{|+\rangle+|-\rangle}{\sqrt{2}}$ and $|1\rangle$ with $\frac{|+\rangle-|-\rangle}{\sqrt{2}}$ gives

$$|\psi_0\rangle=|+\rangle_C\frac{|F_{00}\rangle+|F_{01}\rangle}{\sqrt{2}}+|-\rangle_C\frac{|F_{00}\rangle-|F_{01}\rangle}{\sqrt{2}}. \tag{5}$$

For Eve not being detectable in Step 5, the probability of TP measuring the particle reflected by Charlie in the result $|-\rangle$ should be 0. As a result, we have

$$|F_{00}\rangle=|F_{01}\rangle. \tag{6}$$

On the other hand, assume that Bob's measurement result is $|1\rangle$. After Eve performs $U_F$ on the particles sent back to TP, due to Eq.(3), the state of $B+C+E$ evolves into

$$U_F\left(|10\rangle_{BC}|E_{10}\rangle+|11\rangle_{BC}|E_{11}\rangle\right)=|10\rangle_{BC}|F_{10}\rangle+|11\rangle_{BC}|F_{11}\rangle=|1\rangle_B\left(|0\rangle_C|F_{10}\rangle+|1\rangle_C|F_{11}\rangle\right). \tag{7}$$

Let $|\psi_1\rangle=|0\rangle_C|F_{10}\rangle+|1\rangle_C|F_{11}\rangle$. Replacing $|0\rangle$ with $\frac{|+\rangle+|-\rangle}{\sqrt{2}}$ and $|1\rangle$ with $\frac{|+\rangle-|-\rangle}{\sqrt{2}}$ gives

$$|\psi_1\rangle=|+\rangle_C\frac{|F_{10}\rangle+|F_{11}\rangle}{\sqrt{2}}+|-\rangle_C\frac{|F_{10}\rangle-|F_{11}\rangle}{\sqrt{2}}. \tag{8}$$

For Eve not being detectable in Step 5, the probability of TP measuring the particle reflected by Charlie in the result $|-\rangle$ should be 0. As a result, we have

$$|F_{10}\rangle=|F_{11}\rangle. \tag{9}$$

(iii) Thirdly, consider the case that Bob has chosen to CTRL and Charlie has chosen to SIFT. As a result, the state of $B+C+E$ will be $|00\rangle_{BC}|E_{00}\rangle+|10\rangle_{BC}|E_{10}\rangle$ if Charlie's measurement result is $|0\rangle$ or $|01\rangle_{BC}|E_{01}\rangle+|11\rangle_{BC}|E_{11}\rangle$ if Charlie's measurement result is $|1\rangle$.

Assume that Charlie's measurement result is $|0\rangle$. After Eve performs $U_F$ on the particles sent back to TP, due to Eq.(3), the state of $B+C+E$ evolves into

$$U_F\left(|00\rangle_{BC}|E_{00}\rangle+|10\rangle_{BC}|E_{10}\rangle\right)=|00\rangle_{BC}|F_{00}\rangle+|10\rangle_{BC}|F_{10}\rangle=|0\rangle_C\left(|0\rangle_B|F_{00}\rangle+|1\rangle_B|F_{10}\rangle\right). \tag{10}$$

Let $|\phi_0\rangle=|0\rangle_B|F_{00}\rangle+|1\rangle_B|F_{10}\rangle$. Replacing $|0\rangle$ with $\frac{|+\rangle+|-\rangle}{\sqrt{2}}$ and $|1\rangle$ with $\frac{|+\rangle-|-\rangle}{\sqrt{2}}$ gives

$$|\phi_0\rangle=|+\rangle_B\frac{|F_{00}\rangle+|F_{10}\rangle}{\sqrt{2}}+|-\rangle_B\frac{|F_{00}\rangle-|F_{10}\rangle}{\sqrt{2}}. \tag{11}$$

For Eve not being detectable in Step 5, the probability of TP measuring the particle reflected by Bob in the result $|-\rangle$ should be 0. As a result, we have

$$|F_{00}\rangle=|F_{10}\rangle. \tag{12}$$

On the other hand, assume that Charlie's measurement result is $|1\rangle$. After Eve performs $U_F$ on the particles sent back to TP, due to Eq.(3), the state of $B+C+E$ evolves into

$$U_F\left(|01\rangle_{BC}|E_{01}\rangle+|11\rangle_{BC}|E_{11}\rangle\right)=|01\rangle_{BC}|F_{01}\rangle+|11\rangle_{BC}|F_{11}\rangle=|1\rangle_C\left(|0\rangle_B|F_{01}\rangle+|1\rangle_B|F_{11}\rangle\right). \tag{13}$$

Let $|\phi_1\rangle = |0\rangle_B|F_{01}\rangle + |1\rangle_B|F_{11}\rangle$. Replacing $|0\rangle$ with $\frac{|+\rangle+|-\rangle}{\sqrt{2}}$ and $|1\rangle$ with $\frac{|+\rangle-|-\rangle}{\sqrt{2}}$ gives

$$|\phi_1\rangle = |+\rangle_B \frac{|F_{01}\rangle+|F_{11}\rangle}{\sqrt{2}} + |-\rangle_B \frac{|F_{01}\rangle-|F_{11}\rangle}{\sqrt{2}}. \tag{14}$$

For Eve not being detectable in Step 5, the probability of TP measuring the particle reflected by Bob in the result $|-\rangle$ should be 0. As a result, we have

$$|F_{01}\rangle = |F_{11}\rangle. \tag{15}$$

According to Eqs.(6), (9), (12) and (15), we have

$$|F_{00}\rangle = |F_{01}\rangle = |F_{10}\rangle = |F_{11}\rangle = |F\rangle. \tag{16}$$

(iv) Fourthly, consider the case that both Bob and Charlie have chosen to CRTL. As a result, the state of $B+C+E$ will be $|00\rangle_{BC}|E_{00}\rangle + |01\rangle_{BC}|E_{01}\rangle + |10\rangle_{BC}|E_{10}\rangle + |11\rangle_{BC}|E_{11}\rangle$.

After Eve performs $U_F$ on the particles sent back to TP, due to Eq.(3), the state of $B+C+E$ evolves into

$$U_F\left(|00\rangle_{BC}|E_{00}\rangle + |01\rangle_{BC}|E_{01}\rangle + |10\rangle_{BC}|E_{10}\rangle + |11\rangle_{BC}|E_{11}\rangle\right) =$$
$$|00\rangle_{BC}|F_{00}\rangle + |01\rangle_{BC}|F_{01}\rangle + |10\rangle_{BC}|F_{10}\rangle + |11\rangle_{BC}|F_{11}\rangle. \tag{17}$$

For Eve not being detectable in Step 5, TP should measure the state of $B+C$ in the result $|++\rangle$. After inserting Eq.(16) into Eq.(17), we find out that the following relation is naturally established:

$$U_F\left(|00\rangle_{BC}|E_{00}\rangle + |01\rangle_{BC}|E_{01}\rangle + |10\rangle_{BC}|E_{10}\rangle + |11\rangle_{BC}|E_{11}\rangle\right) = |++\rangle_{BC}|F\rangle. \tag{18}$$

(v) Applying Eq.(16) into Eqs.(3), (4), (7), (10) and (13) produces

$$U_F\left(|x_1 x_2\rangle_{BC}|E_{x_1 x_2}\rangle\right) = |x_1 x_2\rangle_{BC}|F\rangle \text{ for } x_1 x_2 = 00, 01, 10, 11, \tag{19}$$

$$U_F\left(|00\rangle_{BC}|E_{00}\rangle + |01\rangle_{BC}|E_{01}\rangle\right) = |00\rangle_{BC}|F_{00}\rangle + |01\rangle_{BC}|F_{01}\rangle = |0+\rangle_{BC}|F\rangle, \tag{20}$$

$$U_F\left(|10\rangle_{BC}|E_{10}\rangle + |11\rangle_{BC}|E_{11}\rangle\right) = |10\rangle_{BC}|F_{10}\rangle + |11\rangle_{BC}|F_{11}\rangle = |1+\rangle_{BC}|F\rangle, \tag{21}$$

$$U_F\left(|00\rangle_{BC}|E_{00}\rangle + |10\rangle_{BC}|E_{10}\rangle\right) = |00\rangle_{BC}|F_{00}\rangle + |10\rangle_{BC}|F_{10}\rangle = |+0\rangle_{BC}|F\rangle, \tag{22}$$

$$U_F\left(|01\rangle_{BC}|E_{01}\rangle + |11\rangle_{BC}|E_{11}\rangle\right) = |01\rangle_{BC}|F_{01}\rangle + |11\rangle_{BC}|F_{11}\rangle = |+1\rangle_{BC}|F\rangle, \tag{23}$$

respectively.

According to Eqs.(19-23), it can be concluded that for Eve not inducing errors in Steps 5 and 6, the final state of Eve's probe should be independent of Bob and Charlie's measurement results. Therefore, we have completely proved Theorem 1. ∎

In addition, Eve may utilize the round particle transmissions in our protocol to launch the Trojan horse attacks including the invisible photon eavesdropping attack [79] and the delay-photon Trojan horse attack [80-81]. For eliminating the influence of the invisible photon eavesdropping attack, Bob (Charlie) can insert a filter in front of his (her) devices to filter out the photon signal with an illegitimate wavelength before he (she) deals with it [81-82]. For detecting the delay-photon Trojan horse attack, Bob (Charlie) can use a photon number splitter (PNS) to split each sample quantum signal into two pieces and measure the signals after the PNS with proper measuring bases [81-82]. If the multiphoton rate is unreasonably high, this attack will be detected.

In Step 7, Bob (Charlie) publishes $R_B$ ($R_C$) to TP. Obviously, $G_B^i$ ($G_C^i$) is encrypted with $M_B^i$ ($M_C^i$) and $K_{BC}^i$ in this step. However, Eve has no knowledge about $M_B^i$ ($M_C^i$) and $K_{BC}^i$. As a result, even though she may hear of $R_B^i$ ($R_C^i$) from Bob (Charlie), she still cannot obtain $G_B^i$ ($G_C^i$).

In Step 8, TP tells Bob and Charlie the comparison result of $X$ and $Y$. However, it is helpless for Eve to know $G_B^i$ and $G_C^i$.

It can be concluded now that our protocol is secure against an outside eavesdropper.

### 3.2.2 The participant attack

Participant attack always comes from a dishonest participant. It is generally more powerful than the outside attack and should be paid more attention to, as pointed out by Gao *et al.* [83]. There are totally two cases of participant attack in our protocol, namely, the attack from dishonest Bob or Charlie and the attack from semi-honest TP.

**Case 1: the attack from dishonest Bob or Charlie**

In our protocol, the role of Bob is the same as that of Charlie. Without loss of generality, we assume Bob as the dishonest user who tries his best to obtain the SIFT bits of honest Charlie.

Dishonest Bob may launch his attacks on the transmitted particles in Steps 1 and 2. Firstly, we consider the special attacks

Bob may launch. Bob may disturb the particles from TP to Charlie in Step 1 or the particles from Charlie to TP in Step 2. If Bob is clever enough, he will do as follows. (1) When he chooses to CRTL, in order to decrease the probability of being discovered by TP, he does nothing on the particles from TP to Charlie and from Charlie to TP, as there are no SIFT bits. (2) When he chooses to SIFT, in order to obtain Charlie's SIFT bits, he may try the following three schemes: (i) he intercepts the particles from TP to Charlie, measures them in $\sigma_z$ basis, sends the same states he found to Charlie and does nothing on the particles from Charlie to TP; (ii) he does nothing on the particles from TP to Charlie, intercepts the particles from Charlie to TP, measures them in $\sigma_z$ basis and sends the same states he found to TP; (iii) he intercepts the particles from TP to Charlie, sends his prepared fake particles in $\sigma_z$ basis instead of them to Charlie and does nothing on the particles from Charlie to TP. As Charlie's choices of operations are random to Bob, in these schemes, Bob's disturbance behaviors will be discovered by TP. For example, assume that Bob chooses the scheme of (i). As a result, if Charlie chooses to SIFT, Bob will introduce no errors; but if Charlie chooses to CTRL, Bob will be discovered with the probability of 50% . To sum up, the average error rate introduced by Bob when he chooses the scheme of (i) is 25% .

Secondly, like the outside eavesdropper Eve, Bob may launch the entangle-measure attack which can be modeled as two unitaries $U_E$ and $U_F$, where $U_E$ attacks particles as they go from TP to him and Charlie, $U_F$ attacks particles as they go back from him and Charlie to TP, and $U_E$ and $U_F$ share a common probe space with initial state $|0\rangle_E$. As a result, we can directly obtain the following Theorem 2 from Theorem 1 of Ref.[71], which means that although Bob knows his own choices of operations and his own measurement results, he will still get no information on the SIFT bits of Charlie if he escapes from the security checks in Steps 5 and 6.

**Theorem 2.** *Suppose that Bob performs attack $(U_E, U_F)$ on the particles from TP to him and Charlie and back to TP. For this attack inducing no error in Steps 5 and 6, the final state of Bob's probe should be independent of Charlie's measurement results. As a result, Bob gets no information on the SIFT bits of Charlie.*

In Step 7, Bob may hear of $R_C$ from Charlie when she publishes it to TP. However, $M_C^i$ is completely random to Bob. The only thing he can do is to guess its value randomly. Since $G_C^i$ is encrypted with $M_C^i$, he cannot extract $G_C^i$ from $R_C^i$.

In Step 8, Bob hears of the comparison result of $X$ and $Y$ from TP. He cannot obtain $G_C^i$ either.

**Case 2: the attack from semi-honest TP**

In our protocol, TP is allowed to misbehave on his own but cannot conspire with either of Bob and Charlie.

In Step 7, TP hears of $R_B$ ($R_C$) from Bob (Charlie). Although TP knows the values of $M_B$ ($M_C$) through OPERATION 4, he still cannot obtain $G_B^i$ ($G_C^i$) from $R_B^i$ ($R_C^i$), as he has no chance to know $K_{BC}^i$.

It should be emphasized that TP knows the comparison result of $X$ and $Y$ after the calculations of Step 8.

## 4 Discussion and Conclusion

We further compare our protocol with the previous SQPC protocols of Refs.[77-78] in detail. The comparison results are summarized in Table 3. From Table 3, it is apparent that our protocol exceeds the previous SQPC protocols of Refs.[77-78] in both initial prepared quantum resource and quantum measurement for TP. Moreover, our protocol also takes advantage over the protocol of Ref.[77] in the usage of quantum entanglement swapping. However, our protocol is defeated by the protocol of Ref.[77] in the usage of pre-shared SQKD/SQKA key.

Table 3  Comparison of our SQPC protocol and the previous SQPC protocols

|  | The protocol of Ref.[77] | The protocol of Ref.[78] | Our protocol |
| --- | --- | --- | --- |
| Characteristic | Measure-resend | Measure-resend | Measure-resend |
| Initial prepared quantum resource | Bell entangled states | Bell entangled states | Two-particle product states |
| Quantum measurement for TP | Bell state measurements and single-photon measurements | Bell state measurements and single-photon measurements | Single-photon measurements |
| Quantum measurement for users | Single-photon measurements | Single-photon measurements | Single-photon measurements |
| Type of TP | Semi-honest | Semi-honest | Semi-honest |
| TP's knowledge about the comparison result | Yes | Yes | Yes |
| Usage of quantum entanglement swapping | Yes | No | No |
| Usage of pre-shared SQKD/SQKA key | No | Yes | Yes |

With respect to the experimental implementation, our protocol requires the quantum technologies for preparing single photons and two-particle product states, performing $\sigma_Z$ basis and $\sigma_X$ basis measurements and storing single photons. The single photon source [84-85] has been realized in experiment. The measurement of single-photon can be realized through single-photon detector [86]. The storage of single photons can be realized via optical delays in a fiber [8]. It can be concluded that our protocol can be realized with current quantum technologies.

In summary, by using two-particle product states as the initial prepared quantum resource, we propose a SQPC protocol with the measure-resend characteristic for the equality comparison of private secrets from two classical users under the help of a quantum TP. The quantum TP is semi-honest in the sense that he is allowed to misbehave on his own but cannot conspire with either of users. We validate the output correctness and the security against the outside attack and the participant attack in detail. Compared with the previous SQPC protocols, the advantage of our protocol lies in that it only employs two-particle product states as the initial prepared quantum resource, only requires TP to perform single-photon measurements and does not need quantum entanglement swapping. Our protocol is feasible with current quantum technologies.

**Acknowledgments**
Funding by the Natural Science Foundation of Zhejiang Province (Grant No.LY18F020007) is gratefully acknowledged.

**Appendix**

**Lu and Cai's three-party circled SQKD protocol:**

In Lu and Cai's three-party circled SQKD protocol, classical Bob and classical Charlie can establish a sequence of random key bits between them with the assistance of a quantum TP. This protocol is described as follows.

In each run, TP always prepares a qubit in $\sigma_X$ basis and sends it out through the quantum channel. When a qubit is arriving, both Bob and Charlie can either let it go undisturbed or measure it in $\sigma_Z$ basis, prepare a fresh one randomly in $\sigma_Z$ basis and send it. TP receives the travel back qubit and measures it in $\sigma_X$ basis or $\sigma_Z$ basis randomly. There are three possibilities: (p0) Neither Bob nor Charlie disturbed the travel qubit; (p1) One of them measured the travel qubit but the other did not; (p2) Both Bob and Charlie measured the travel qubit.

After all of TP's qubits have been distributed, Bob, Charlie and TP can publish their operations through their classical channels. When (p0) happened, the travel qubit has not been disturbed so that they can use this run as CTRL if TP has measured it in $\sigma_X$ basis. When (p1) happened, Bob or Charlie publishes the state of the travel back qubit and then they can also use this run as CTRL if TP has measured the travel qubit in $\sigma_Z$ basis. When (p2) happened, Charlie knows the state Bob prepared so that Bob and Charlie share a common bit as SIFT bit. In the end, Bob and Charlie will publish some of their SIFT bits to verify QBER in SIFT. If both QBER in SIFT and QBER in CTRL are tolerable, Bob and Charlie can use the rest SIFT bits as INFO bits to generate final key bits after error-correction and privacy amplification.

It is necessary to emphasize that TP cannot share Bob and Charlie's key bits since Charlie refreshes each travel qubit in SIFT.